\documentclass{elsart5p}

\usepackage{epsfig}
\usepackage{amssymb}

\begin{document}

\begin{frontmatter}



\title{An orientable time of flight detector for cosmic rays}


\author[Rome]{M. Iori}
\author[Rome]{and A. Sergi\thanksref{Perugia}}
\thanks[Perugia]{now at INFN-Perugia, Italy}
\address[Rome]{``Sapienza'' University of Rome, Piazzale A. Moro 5 00185 Rome, Italy}
\begin{abstract}
Cosmic ray studies, in particular UHECR, can be in general supported by a directional, easy deployable,
simple and robust detector. The design of this detector is based on the time of flight between two parallel
tiles of scintillator, to distinguish particle passing through in opposite directions; by fine time resolution
and pretty adjustable acceptance it is possible to select upward(left)/downward(right) cosmic rays. It has
been developed for an array of detectors to measure upward $\tau$ from Earth-Skimming neutrino events with
energy above $10^8~GeV$. The properties and performances of the detector are discussed. Test results from
a high noise environment are presented.
\end{abstract}

\end{frontmatter}

% main text
\section{Introduction}
The interest in Ultra High Energy Cosmic Rays (UHECR) produced a large variety of experiments, with different
purposes and based on several techniques (Cherenkov, air fluorescence and radio waves);
while timing information is often
used to obtain directional information, none of the present techniques is
based on upward/downward discrimination by time of flight. The prototype
detector described here is capable to measure large zenith angle cosmic rays 
as well  to be an element of an orientable surface array of detectors
to measure the signature of Ultra High Energy $\tau$ 
neutrinos using the  Earth skimming strategy \cite{fargion,feng,beacom,zas}, as 
shown in the TAUWER proposal \cite{tauwer}.

\section{Description of the detector}
\begin{figure}[htbp]
\begin{center}
\includegraphics[width=7.cm,totalheight=6.cm]{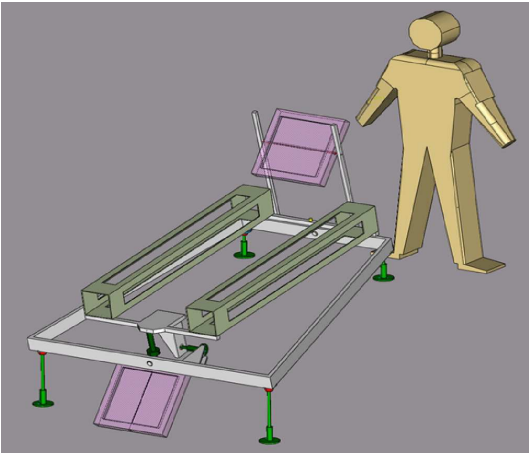}
\caption{\label{fig:modulo1} Schematic view of the detector. The electronic box and the
wireless connection are not shown. }
\end{center}
\end{figure}

\begin{figure}[htbp]
\begin{center}
\includegraphics[width=7.cm,totalheight=7.5cm]{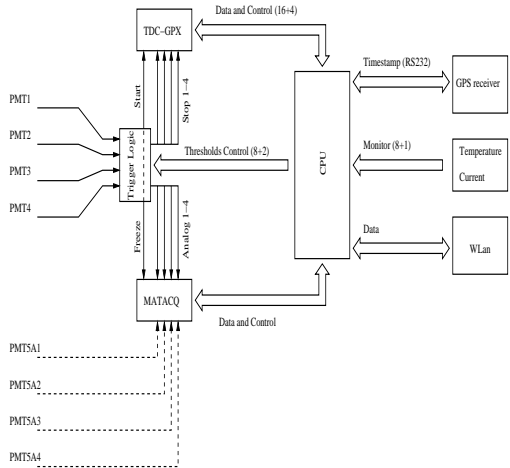}
\caption{\label{fig:board} Hardware block diagram of the DAQ board. }
\end{center}
\end{figure}

Besides the deployment mechanics, not discussed here, the structure of the detector
is very robust and simple, as shown in Fig. \ref{fig:modulo1}; the base detector consists in 
two parallel scintillating plates ($20~\times~20~cm^2$, 1.4 cm thick), separated by $160~cm$, read
by one low voltage R5783 Hamamatsu photomultiplier (PMT), extensively used in CDF muon detector \cite{cdf}. 
Each scintillating plate
is embedded in a PVC box which also contains the PMT. The two boxes are attached
to a metal structure that defines the covered solid angle of about 1.4 $10^{-2}$rad.
The choice of this particular model of PMT is due to its low time resolution 
($\approx 300~ps$) and the possibility to use an autonomous
low voltage power supply, like a solar panel or a wind turbine, to make it
an affordable elementary module of a large area array. 
A custom electronic
board for time and charge analysis, in substitution of standard NIM-CAMAC modules, 
is under development (Fig. \ref{fig:board}).

Thinking about using
this detector for extensive air showers, it would be interesting to have the
possibility to study the shower front structure; this can be done by using
a sampling ADC based on MATACQ, $2.5 ~\mu s$ at $1GS/s$ \cite{Matacq}, that can be used also 
at trigger level
to define the direction of the track. The time of flight is provided by a 
TDC-GPX, by ACAM, with 40 ps or 81 ps resolution \cite{acam}. 
In this case it has to be noticed that the board has to be equipped
with a proper pipeline, to retain raw data until an exchange of trigger
masks of the modules with the central DAQ, by WLAN, provides a global
trigger decision; this is necessary to have the capability, given the data rate
transmission limit, to store the full MATACQ response in relation to events that
involve only single scintillators spread over the whole array.

\subsection{Acceptance}
The present detector was designed to recognize single near horizontal moving
particles and it is basically a narrow angle scintillator counter, illustrated in
Fig. \ref{fig:modulo1}. 
The single module, called \emph{tower}, has a zenith angle range of $\pm~7.5^o$
around the axis and the geometrical acceptance results to be $5.1~cm^2 sr$.
By the time resolution of the order of $1~ns$, it is possible to reject
vertical air showers without need of any shielding and 
to select upward and downward particles passing through the detector
with negligible intrinsic contamination; by coupling
%\begin{figure}[htbp]
%%\begin{center}
%\includegraphics[width=6.cm,totalheight=4.cm]{modulo1.eps}
%\caption{\label{fig:modulo1} Schematic view of the detector.}
%%\end{center}
%\end{fig  by coupling
two towers the acceptance can be  enhanced up to $\pm~ 20^o$ along the azimuthal angle  
if the \emph{towers} are separated of $\approx 60~cm$, 
hence the covered solid angle increases almost by a factor 3. This is
particularly important in situations like a large array whose target are rare
events (i.e. UHE neutrino flux).
\section{Performances}
\begin{figure}[htbp]
\begin{center}
\includegraphics[width=7.cm,totalheight=6.cm]{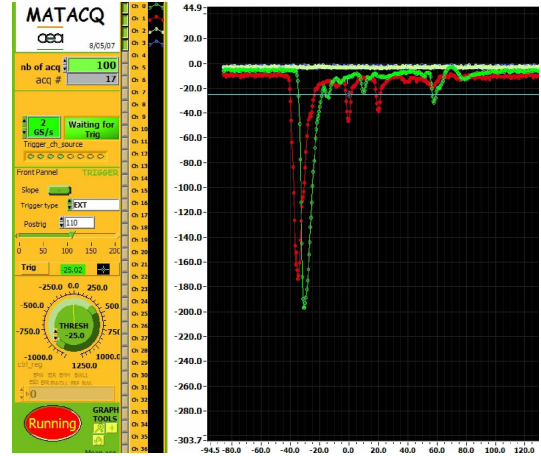}
\caption{\label{fig:Mat} Two photomultiplier signals displayed by Matacq board. 
The units are $mV$ (vertical axis) and $ns$ (horizontal axis). }
\end{center}
\end{figure}

To optimize the performances of the detector several tests were performed
with different tile sizes at the High Altitude Research Station Jungfraujoch (HFSJG), 
located in Switzerland
at $\approx 3600~m$ above the sea level. In some tests the data were collected by the 
MATACQ board only (Fig. \ref{fig:Mat}).
The detector has shown a good upward/downward discrimination capability in all our
tests. 
In order to reach upward-downward separation requirement the light collection
technique has been optimized focusing on the time resolution, disfavoring the energy
resolution. 
%To obtain this result we had to neglect its energy resolution because
%of the light collection technique, imposed by time resolution requirements.
The definition of a vertical MIP is made by calibration on vertical
downward cosmic rays, to set proper charge cuts to obtain a good time
resolution, slightly reducing the effective area.

\begin{figure}[htbp]
\begin{center}
\includegraphics[width=7.cm,totalheight=6.cm]{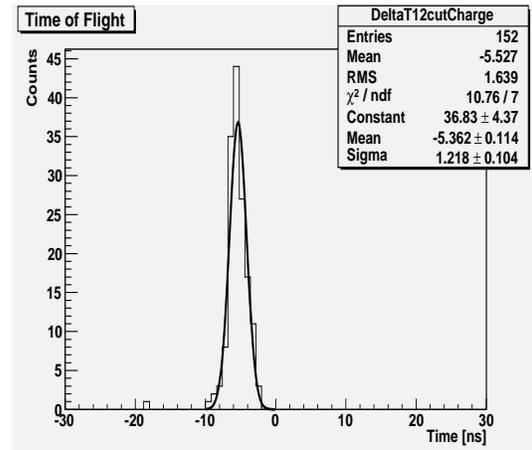}
\caption{\label{fig:tres} The distribution of
time of flight between the two 20x20 $cm^{2}$ tiles $160~cm$ apart,
for downward vertical cosmic ray events.}
\end{center}
\end{figure}

\subsection{Time resolution}
\begin{figure}[htbp]
\begin{center}
\includegraphics[width=7.cm,totalheight=7.cm]{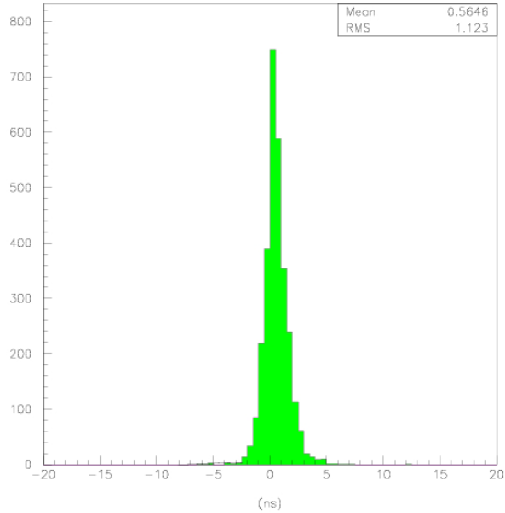}
\caption{\label{fig:2cm} Time of flight between two horizontal overlapped tiles.}
\end{center}
\end{figure}

The time resolution ($\approx 1~ns$), near the PMT  transit time spread, is obtained by
avoiding any reflection in the light collection; the $1~cm^2$ PMT window is directly
coupled, by a silicon pad, to the scintillator, wrapped with Tyvek. This configuration
let the first light arriving at the PMT window dominate for the leading edge
of the signal; it has been chosen after several tests using WLS or clear fibers,
which lead to the conclusion that any kind of reflection introduces a stochastic
fluctuation that enlarges the time spread by a factor 3 or 4, or, similarly,
for a random collection point for WLS fibers. Figs. \ref{fig:tres} and \ref{fig:2cm} show the 
time of flight between two 
tiles 160 cm apart and two horizontal overlapped tiles respectively. Both distribution have
an RMS of $\approx 1~ns$.

%\begin{figure}[htbp]
%\begin{center}
%\includegraphics[width=5.cm,totalheight=4.cm]{sig2.eps}
%\caption{\label{fig:matacq} Signal from matacq board.}
%\end{center}
%\end{figure}
\subsection{Upward and downward track discrimination }

\begin{figure}[htbp]
\begin{center}
\includegraphics[width=0.4\textwidth]{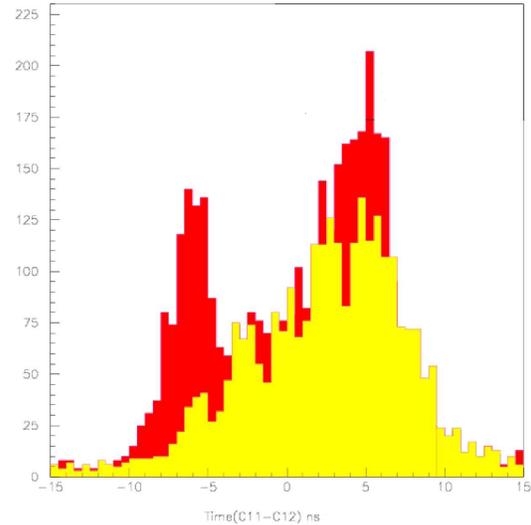}
\caption{\label{fig:pbnopb} Time-of-flight difference between two tiles $\Delta t _{12}$ in a tower
pointing at a zenith angle $95^\circ$ when a lead block of 3-cm thickness is placed in front of one tile 
(yellow shade) and no lead is present (red shade).
The two peaks at -5~ns and +5~ns correspond to up-going and down-going tracks, respectively.
The peaks at -3, 0 and 3 ns are due to parallel tracks, most likely vertical,
where one of them hits an ``inefficient region'', i.e. a corner, of one of the tiles.
}
\end{center}
\end{figure}

The environment in which the detector is supposed to work could present contamination
of the time signal by simulating the right time difference by events different
from particles crossing the detector. It gives the possibility to access to a higher 
flux of "not interesting" cosmic rays, in particular secondary photons.
It was previously understood, by a test with a radioactive source ($^{90}Sr$), that the
region of scintillator outside the field of view of the PMT window is disfavored for
timing and signal amplitude, because it needs at least one reflection to reach the
photocathode, resulting in a hit delayed of about $3~ns$. In the tested prototype
this effect can be observed when it is set exactly horizontal; in this geometrical
condition the vertical, dominant, component of CRs has about $50\%$ probability
to hit an efficient or inefficient region of the side of the scintillator. This means
that, in presence of events of two time correlated and 0 delayed particles, like pair
production from a vertical photon, we expect a structured time signal contamination,
besides the signal ($\approx -5,5~ns$), consisting in 3 peaks ($\approx -3,0,3~ns$), 
due to the combined probability of detection by an efficient/inefficient region 
of the scintillator.
This effect changes with zenith angle orientation drifting with $\sin\theta$, resulting
in the structure showed by Fig. \ref{fig:pbnopb}.

\begin{figure}[htbp]
\begin{center}
\includegraphics[width=0.5\textwidth]{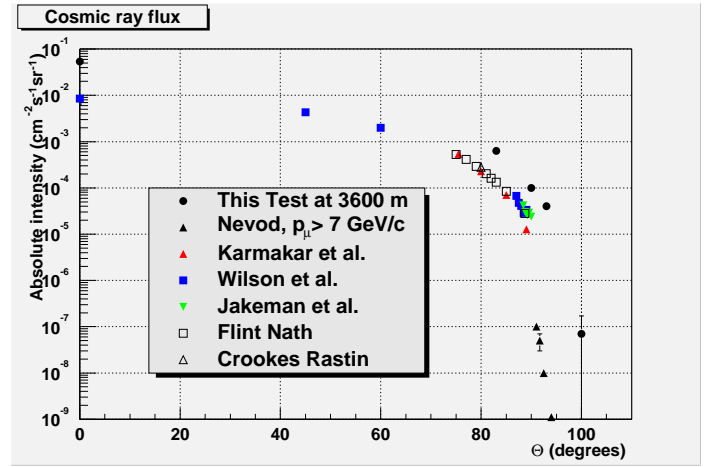}
\caption{\label{fig:flux} Measurements of fluxes at several zenith angles in the 
High Altitude Research Station, Jungfraujoch  compared with results from other experiments at sea level,
(8;9;10;11;12;13).}
\end{center}
\end{figure}

Tests were performed on the possibility that multiple reflections of the emitted light on the surface and on
the border of the tile create peaks on TOF distribution,
to understand the contribution from the surface and the border. 
These results suggest that a blackened border reduce the multiple reflection.
Further test are in progress.
\subsection{Measurement of cosmic ray flux at 3600 m }

Fig. \ref{fig:flux} shows the measurements of cosmic ray flux in the 
High Altitude Research Station Jungfraujoch in a zenith interval of 80$^{o}$ to 100$^{o}$ and at 0$^{o}$.

%\begin{figure}[!htbp]
%\begin{center}
%\includegraphics[width=0.6\textwidth]{fluxes.eps}
%\caption{\label{fig:flux} Measurements of fluxes at several zenith angles in the
%High Altitude Jungfraujoch Station compared with results from other experiments at sea level,
%\cite{Nevo,Karma,Wilson,Jake,Flint,Crookes}.}
%\end{center}
%\end{figure}

\section{Summary}
The detector shown can distinguish upward and downward particles from cosmic rays and
collect data within time interval of $2.5~ \mu s$ at $1GS/s$. By its 
mechanical structure it is possible to change easily the orientation of the detector. In
particular this feature, joint with its intrinsic directionality, make it feasible
for an orientable array for UHECRs; this opens the possibility to use the expertise
on quasi-vertical showers on quasi-horizontal physics, as it is for the TAUWER
proposal.

\section{Acknowledgements:}
To the Director of the International Foundation
High Altitude Research Stations Jungfraujoch and
Gornergrat, Prof. Erwin  Fl\"uckiger, to  Dr. Rolf B\"utikofer, to 
the Technical Staff working in the Jungfraujoch Station, to Dr. E. Delagnes
(CEA-Saclay) and Dr. D. Breton (IN2P3-Orsay) for the modification of the MATACQ board, 
to Maurizio Perciballi for the mechanical work, to Prof. Jim Russ, to Prof. D. Fargion 
for very useful discussions 
and to the Organizer of the Conference, Prof. A. Capone.


\begin{thebibliography}{00}

\bibitem{fargion} D.Fargion et al. Astrophys. J. 613 (2004) 1285. 
\bibitem{feng} J.L. Feng et al. Phys. Rev. Lett. 88 (2002) 161102.
\bibitem{beacom} J.F. Beacom et al. Phys. Rev. D 68 (2003) 093005.
\bibitem{zas} E. Zas, New J.Phys.7:130,2005.
%\bibitem{orig} M. Iori, A. Sergi and D. Fargion, astro-ph/0409159.
\bibitem{tauwer} M. Iori et al. arXiv:astro-ph/0602108.
\bibitem{cdf} A. Artikov et al. Nucl.Instrum.Meth.A538:358-371,2005.
\bibitem{Matacq} E. Delagnes, D. Breton, Echantillonneur analogique rapide grande
profondeur memoire, French patent n01-05607 April 26th 2001. US patent 6,859,375 
Feb 22nd 2005: fast analog sampler with great memory depth. 
\bibitem{acam} M. Mota et al. IEEE Journal of Solid State Circuits Vol 34 (1999) 1360.
\bibitem{Nevo} Nevod Collaboration 29th Int Conference Pune 2005 101-104.
\bibitem{Karma} N.L. Karmakar A. Paul and N. Chaudhuri, Nuovo Cimento B17 (1973) 173.
\bibitem{Wilson} B.C. Wilson, Can. J. Phys. 37 (1959) 19.
\bibitem{Jake} D. Jakeman et al. Can J. Phys. 34 (1956) 432.
\bibitem{Flint} R.W. Flint, R.B. Hicks and S. Standil, Can J. Phys. 50 (1972) 843.
\bibitem{Crookes} J.N. Crookes and B.C. Rastin, Nucl. Phys. B39 (1972) 493.

\end{thebibliography}
\end{document}